\documentclass[a4paper,10pt,conference]{ieeeconf}

\usepackage{times}
\usepackage[linesnumbered,lined]{algorithm2e}
\usepackage{amsmath}
\usepackage{latexsym}
\usepackage{theorem}
\usepackage{epsfig}
\usepackage{subfigure}
\usepackage{setspace}
\usepackage{cite}
\usepackage{flushend}
\usepackage[absolute,overlay]{textpos}

\begin{document}
\title{Instruction-set Selection for Multi-application based ASIP Design: An Instruction-level Study}

\author{\authorblockN{Roshan Ragel\authorrefmark{1},
Swarnalatha Radhakrishnan\authorrefmark{1}, and
Angelo Ambrose\authorrefmark{2}}
\authorblockA{\authorrefmark{1} Department of Computer Engineering  \\
 University of Peradeniya, Sri Lanka}
\authorblockA{\authorrefmark{2} School of Computer Science and Engineering\\
 University of New South Wales, Australia}}

\maketitle

\begin{abstract}
Efficiency in embedded systems is paramount to achieve high performance while consuming less area and power.
Processors in embedded systems have to be designed carefully to achieve such design constraints. 
Application Specific Instruction set Processors (ASIPs) exploit the nature of applications to
design an optimal instruction set. Despite being not general to execute any application,
ASIPs are highly preferred in the embedded systems industry where the devices are produced to satisfy a certain type of 
application domain/s (either intra-domain or inter-domain). Typically, ASIPs are designed from a base-processor 
and functionalities are added for applications.  This paper studies the multi-application ASIPs 
and their instruction sets, extensively analysing the instructions for inter-domain and intra-domain
designs. Metrics analysed are the reusable instructions and the extra cost to add a certain application. 
A wide range of applications from various application benchmarks (\emph{MiBench}, \emph{MediaBench} and 
\emph{SPEC2006}) and domains are analysed for two different architectures 
(\emph{ARM-Thumb} and \emph{PISA}). Our study shows that the intra-domain applications contain larger number of common instructions,
whereas the inter-domain applications have very less common instructions, regardless of the architecture (and therefore the ISA).
\end{abstract}

\section{Introduction} \label{introduction}
\begin{figure}[ht!]
\centering
\includegraphics[width=6.5cm]{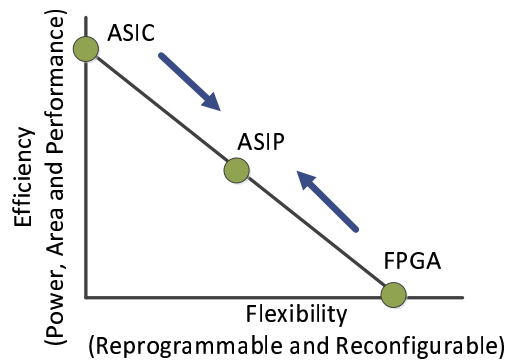}
\caption{ASIC vs. ASIP vs. FPGA}
\label{asipintro}
\end{figure}

Embedded systems are the realm in current civilisation and their omni-presence in modern technology in the form of 
mobile phones, network devices, computers, medical devices, automotive and other applications is obvious.  
Power and energy consumption, device size, durability and reliability are some of the major 
properties which are expected from such embedded devices. Hence it is imperative that the
embedded systems be optimised for the needs of its application to achieve maximum efficiency.
Application Specific Integrated Circuits (ASICs) are specifically made in hardware to execute a functionality
with extremely efficient power, area and performance budgets.  Despite being heavily used in the industry for System-in-Chip designs,
ASICs are hardly flexible and can not be reused for a different type of application. Field Programmable Gate Arrays (FPGA), on the other hand,
are highly flexible, but  inefficient for power and performance. FPGAs are still considered as prototyping platforms for embedded systems mainly due to their
inefficiency in area and power. 
Figure~\ref{asipintro} depicts an illustrative diagram of different technologies 
including ASIC and FPGA. As shown, 
the ASIC is  efficient but lacks flexibility, whereas the FPGAs are flexible, nevertheless costs performance, power and area. To hit a reasonable balance between
ASICs and FPGAs, Application Specific Instruction-set Processors (ASIPs) are considered as the appropriate choice. As shown in Figure~\ref{asipintro},
ASIPs, which are the latest technological trend in embedded systems \cite{glkler10designof}, are conceived by tightening up the flexibility from FPGAs and releasing the efficiency from ASICs. 

ASIPs are formed using a hardware/software co-design process, where instructions are chosen for a processor, based on the behaviour of the application to be deployed. Applications are designed using instructions which are then executed in a processor.   
Such a technology improves the design productivity due to the simplicity in the software implementation process.  Furthermore, the hardware/software co-design approach 
improves productivity by allowing the hardware to be reused and reprogrammed \cite{glkler10designof}. The complexity of the design is reduced, thus Non-Recurring 
Engineering (NRE) cost is also decreased. The instruction set of an ASIP is tailored to benefit a specific application or a known set of applications. 
Such an instruction set based solution provides high degree of flexibility, supporting yet-to-be introduced standards 
and implementations. This provides reasonable tolerance in design changes which might arise in the future. 
ASIPs are typically modelled using high level languages \cite{glkler10designof}, which allows a relatively easy
and methodical approach to design applications on a resource stringent hardware. 

Instructions in an ASIP is an integral component to decide the functionality and its efficiency. 
Since complex application programs use hundreds of types of processor instructions, selecting and designing most suitable 
instructions to achieve the highest performance in an optimised way is a major challenge in the design process of an ASIP.
An instruction set (also known as the instruction set architecture, ISA) serves as the interface 
between hardware and software in a computer system. In an application specific environment, the system 
performance can be improved by designing an instruction set that matches the characteristics of the hardware 
and the application \cite{huang95synthesis}. 

From a cost and performance perspective, types of instructions 
used in given applications is vital. Approaches to instruction set generation 
for an ASIP can be classified as either instruction set synthesis approach \cite{gschwind99instruction,huang95synthesis} 
 or instruction set selection approach \cite{alomary93anasip,imai96anew,liem94instruction,shu96instructionse}  
on the basis of how the instructions are generated. In synthesis approach, instruction set 
is synthesised for a particular application based on the application requirements, while in selection 
approach, a superset of instructions is available and a subset of them is selected to satisfy the 
performance requirements within the architectural constraints \cite{jain01asipdesign}. 

Instructions in an application are affected by three factors: 1), functionality of the application and its 
relationship to the ISA; 2), behaviour of the compiler's code generation; and, 3), coding style. We only
focus on the first one which is the most critical of all. In this paper we perform an 
instruction-level study to realise the nature of instructions used within application domains (i.e., intra-domain) and across application 
domains (i.e., inter-domain). Such a study allows us to envision the effect on instruction commonalities and uniqueness for application 
specific instruction sets. This paper provides an insight into the instruction usage in applications to evaluate 
the intra-domain and inter-domain costs involved for integration.    

\section{Motivation} \label{motivation}
\begin{figure}[ht!]
\centering
\includegraphics[width=7cm]{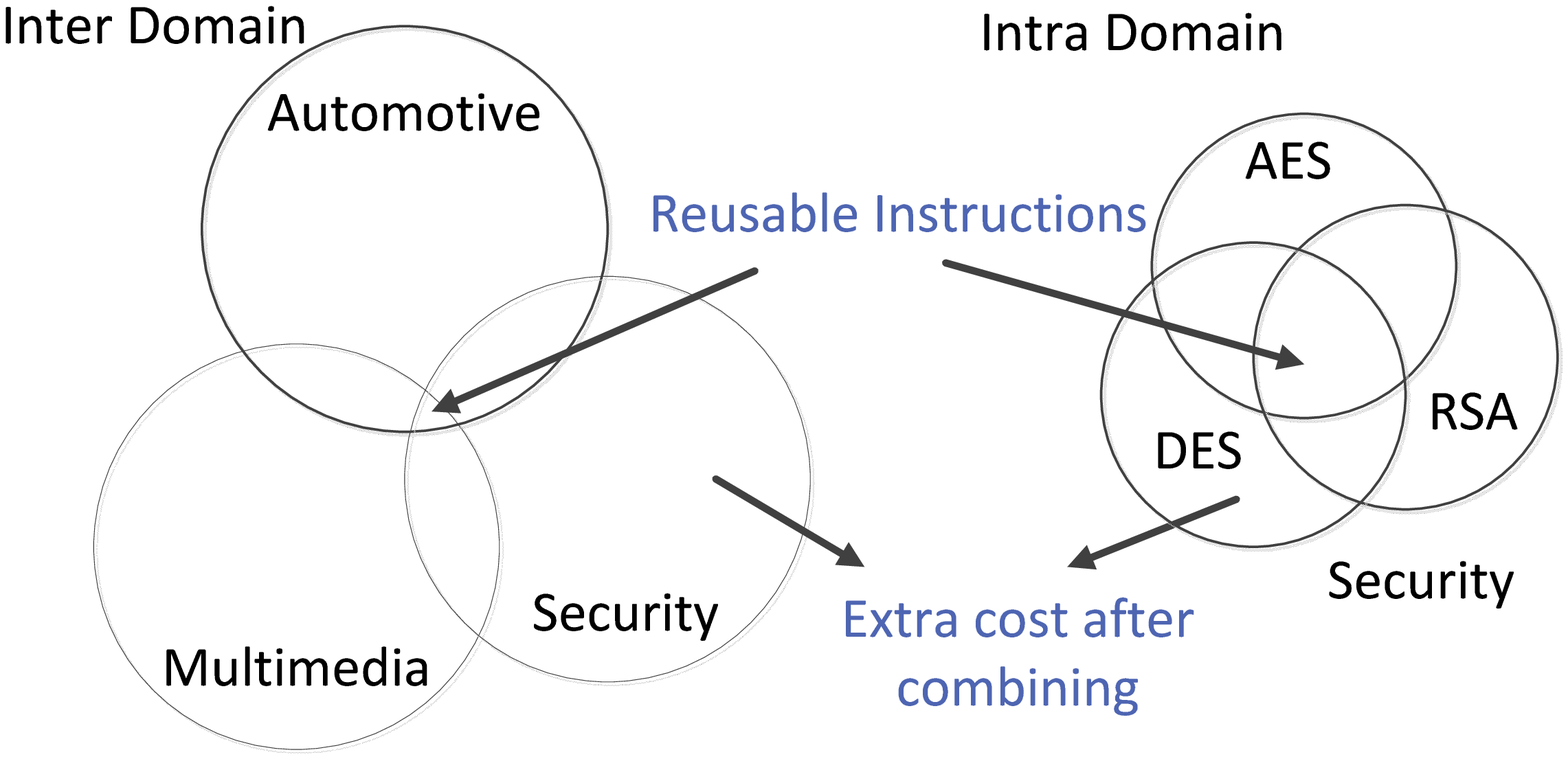}
\caption{Application Domains, Applications and Instructions}
\label{asipisaintro}
\end{figure}

The efficacy of ASIP applications depends on the optimal use of the instructions. Figure~\ref{asipisaintro} 
illustrates three different application domains; automotive, multimedia and security. The Security domain
is illustrated with three applications: AES, DES and RSA, combined in intra-domain.  
If we are to design 
an embedded system to include applications from these three domains (i.e., inter-domain), it is necessary to realise the extra 
cost involved for integration in terms of instructions which is directly related to the design time and effort. 
Since the application domains have quite a significant functional difference, we expect to find very less 
commonality (very less reusable instructions) in the instructions across the three domains. The applications inside an application domain (i.e.,
intra-domain) are 
expected to contain much less uniqueness in instructions (less additional cost and high reusability\footnote{we refer 
to this intersecting instructions as reusable instructions
which are typically built as base processors in state-of-the-art ASIPs \cite{xtensa}}) across difference applications, due to similar type 
of operations (i.e., functionalities) being performed. We endeavour to validate this hypothesis by studying 
the instructions being used in inter-domain and intra-domain applications. It is further important to 
evaluate the contribution of the instruction set to the ASIP design, compared to the coding style and the 
nature of the compiler.

Embedded systems are evolving rapidly and the amount of applications 
executed in an embedded system (such as mobile phones) range in the order of tens to hundreds, while still growing. 
Hence it is necessary for the designer to realise the additional cost involved in integrating applications and 
the  means of reusability to improve the design process. We evaluate these two properties: 1), extra cost and 2), reusability,
at the instruction level.

\section{Related Work} \label{relatedwork}
ASIP systems have become the norm in embedded systems to achieve high performance while being able to consume 
low power \cite{algochip, glkler10designof}. Selection of instructions for an ASIP has been widely studied.
We discuss the most appropriate studies in this section.    

One approach for generating instruction sets is by considering the datapath model. In 1994, Praet et al. 
\cite{vanpraet94instructionset} have shown how instruction selection for ASIPs can be performed by 
generating a combined instruction 
set and datapath model from the instruction set. Operation bundling was performed on the model with an abstract
datapath. This methodology still requires refinement and testing.
Then Kucukcakar \cite{kucukcakar99anasip} came up with an architecture and a co-design methodology to improve the performance of embedded 
system applications through instruction-set customisation, based on a similar kind of concept. Although 
these methodologies improve the performance of ASIPs they failed to consider the design constraints such as area, 
power consumption, NRE cost etc.

\begin{table*}[ht!]
	\centering
	\scriptsize
	\caption{Benchmark Applications Used}
	\label{tab:BMUsed}
	\begin{tabular}{|l|l|l|l|l|l|} 					\hline
	  \multicolumn{6}{|c|}{\bf Applications per Domains} \\ \hline
		Automotive (AM) & Office (OF) & Security (SE) & Telecomm (TC) & MediaBench (MB) & Spec.CPU2006Int (SP) \\	\hline
		BasicMath       & GhostScript & Blowfish      & Adpcm         & Epic              & BZip2              \\ \hline
		BitCount        & ISpell      & PGP           & CRC32         & G721              & MCF                \\ \hline
		QuickSort       & RSynth      & Rijndael      & FFT           & H263enc           & Hmmer              \\ \hline
		Susan		   & StringSearch& Sha           & GSM           & MPEG2enc          & Sjeng               \\ \hline
	\end{tabular}
\end{table*}

One of the early work on methodologies to maximise performance of ASIP under design constraints, such as area,
power consumption and NRE cost, is \cite{cheung03rapidconf}. The authors in \cite{cheung03rapidconf} proposed a rapid
instruction selection approach from their library of 
pre-designed specific instructions to be mapped on a set of pre-fabricated co-processors/functional units .  
As a result, the authors  in \cite{cheung03rapidconf} were able to significantly
 increase application performance while satisfying area constraints. This methodology uses a combination of 
simulation, estimation and a pre-characterised library of instructions, to select the appropriate co-processors 
and instructions. Alomary et al. \cite{alomary93anasip} proposed a new formalisation 
and an algorithm that considers the functional module sharing. This method allows designers to predict 
the performance of their designs before implementation, which is an important feature for producing a 
high quality design in reasonable time. In addition to that, an efficient algorithm for automatic selection 
of new application-specific instructions under hardware resources constraints is introduced in \cite{galuzzi06automaticselection}. 
The main drawback of this algorithm is the un-optimised  
Very-High-Speed Integrated circuits Hardware Description Language (VHDL) model.

Researchers have already proposed automated techniques in ASIP design process to achieve best performance under
certain design constraints. Almer et al. \cite{almer09anendtoend} presented a complete tool-chain for automated 
instruction set extension, micro-architecture optimisation and complex instruction selection, based on GCC compiler.
Huang and Despain in \cite{huang94generating} proposed a single formulation, combining the 
problem of instruction set design, micro-architecture design and instruction set mapping. The formulation 
receives as inputs the application, architecture template, objective function and design constraints, and generates as 
outputs the instruction set for the application. Similarly, Zhu et al. \cite{zhu06afast} presented
a design automation approach, referred to as Automatic Synthesis of Instruction-set Architectures (ASIA),
to synthesise instruction sets from application benchmarks. The problem of designing instruction sets
was formulated as a modified scheduling problem in \cite{zhu06afast} .
In \cite{wu08memorymodels}, a design flow was proposed to automatically generate 
Application-Specific Instructions (ASIs) to improve performance with memory access considerations. 
The ASIs are selected not only based on the instruction latency but also the memory access.

Once the instructions are chosen for an ASIP, the selected instructions are evaluated. 
Authors in \cite{huang95synthesis} and \cite{liem94instruction} introduced methods
to evaluate instruction sets with several design constraints. Peymandoust et al. \cite{peymandoust03automatic}
automatically grouped and evaluated data-flow operations in the application as potential custom instructions. 
A symbolic algebra approach is utilised to generate the custom instructions with high level arithmetic optimisations. 

Considering the process of instruction selection and evaluation for ASIPs, in this paper, we
perform an application  analysis (at the instruction level) to identify the commonalities and 
uniqueness for intra-domain and inter-domain applications. We evaluate the applications based on the 
extra cost for application integration and reusability of common instructions. Such an analysis will enlighten the designer in performing
smart instruction selection. 

\section{Methodology} \label{methodology}
As highlighted in Section~\ref{motivation}, our objective is to study the reusability and extra cost  of  multi-application
ASIPs (named mASIPs) in terms of instruction set utilisation. The method we device to perform this study is described in this section. For 
every  instance  of  our  experiment,  we  choose a set of target applications, one or many of which can be deployed in our
mASIP. Therefore, the target application set is a list of potential applications for an mASIP design. 
The target applications can come from a single application domain (such applications are identified
as intra-domain applications) or multiple application domains (such applications are identified as inter-domain applications).

With respect to the instruction set design, the mASIPs can be built using two
phases: one, designing and building a \emph{base processor} with the instruction set necessary for all the applications of the 
target set and two, extending the base processor to cater the rest of the instruction types for the applications to be deployed for a 
particular mASIP. It is worth to note that the applications that will be deployed for a particular mASIP is a subset
of the applications in the target set. We will define the instruction set of the base processor as the \emph{base instruction set}.
Therefore, the \emph{base instruction set} is the set of instructions that are common to all the applications in our
target application set.

With this background, in our study, we calculate the reusability and the extra cost of an mASIP by using the \emph{base 
instruction set} and the rest of the instruction set necessary for building an mASIP. That is, a larger base instruction
set will indicate a higher reusability and a larger additional instructions in phase two of our design would 
indicate a higher extra cost. Both reusability and extra cost of a particular mASIP design will be quantified by the
number of instructions in the base instruction set and the rest of the instructions needed to complete the mASIP design.

\begin{figure}[ht!]
\centering
\includegraphics[width=8cm]{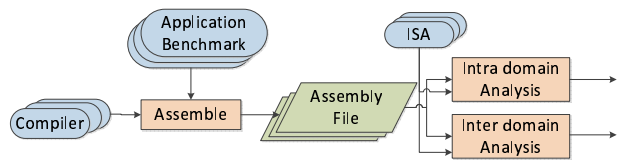}
\caption{Experimental Flow}
\label{exp}
\end{figure}

\section{Experimental Setup} \label{experimental}
Figure~\ref{exp} explains the experimental flow of our study. Application benchmarks are assembled, using two different cross compilers
(targeting two well known instruction set architectures, \emph{ARM-Thumb} and \emph{PISA}), to create the assembly files, indicating all the instructions used. 
It is worth to note that \emph{ARM-Thumb} has 78 instruction in its ISA and \emph{PISA} has 72 instructions (integer instructions only)  in its ISA. We call
them the complete instruction sets for \emph{ARM-Thumb} and \emph{PISA}.
Applications are collected from different domains, four applications each. The assembly files are then
analysed for intra-domain and inter-domain instruction level dependencies. The complete set of ISA of each architecture is another input to the analysis.

\begin{figure*}[th!]
\begin{equation}\label{eq1}
Reusability\,Factor = \frac{(\#\,of\,instructions)_{Base\,ISA} * 100}{(\#\,of\,instructions)_{mASIP}}
\end{equation}
\end{figure*}

\begin{figure*}[th!]
\begin{equation}\label{eq2}
Extra\,Cost\,Factor = \frac{[(\#\,of\,instructions)_{Apps/Domain} -  (\#\,of\,instructions)_{Base\,ISA}] * 100}{(\#\,of\,instructions)_{mASIP}}
\end{equation}
\end{figure*}

\begin{table}[ht!]
	\scriptsize
	\centering
	\caption{Instruction-set Selection for Intra-Domain Applications}
	\label{tab:IntraDomain}
	\begin{tabular}{|l|l|c|c|c|c|c|c|} 					\hline
				    & 			& \multicolumn{6}{|c|}{\bf Number of Instructions} \\ \cline{3-8}
	{\bf Group} 	& {\bf Domain}	& \multicolumn{3}{|c|}{\bf ARM-Thumb ISA}& \multicolumn{3}{|c|}{\bf PISA}\\ \cline{3-8} 
					&			&	Indiv.	 		& 	Inter.		& Union 			&	Indiv.		& Inter.			& Union		\\	\hline
					& BasicMath	&		33 			&				&				&		25		&				&			\\
		AM		 	& BitCount	&		46			&	23			& 49				&		31		&	16			&	40		\\
					& QSort		&		25			&				&				&		19		&				&			\\
					& Susan		&		45			&				&				&		34		&				&			\\	\hline
					
					& GhostScript&		52 			&				&				&		44		&				&			\\
		OF		 	& ISpell		&		29			&	27			& 55				&		50		&	27			&	51		\\
					& RSynth		&		52			&				&				&		40		&				&			\\
					& StringSearch&		40			&				&				&		27		&				&			\\	\hline

					& BlowFish	&		49 			&				&				&		30		&				&			\\
		SE		 	& PGP		&		57			&	30			& 57				&		52		&	22			&	52		\\
					& Rijndael	&		36			&				&				&		30		&				&			\\
					& Sha		&		40			&				&				&		29		&				&			\\	\hline

					& Adpcm		&		39 			&				&				&		32		&				&			\\
		TC		 	& CRC32		&		36			&   25			& 55				&		22		&	16			&	45		\\
					& FFT		&		41			&				&				&		30		&				&			\\
					& GSM		&		54			&				&				&		41		&				&			\\	\hline

					& Epic		&		56 			&				&				&		44		&				&			\\
		MB		 	& G721		&		49			&	45			& 56				&		41		&	35			&	50		\\
					& MPEG2		&		51			&				&				&		43		&				&			\\
					& Rasta		&		53			&				&				&		44		&				&			\\	\hline

					& BZip2		&		57 			&				&				&		50		&				&			\\
		SP	 		& Hmmer		&		45			&	45			& 58				&		29		&	29			&	52		\\
					& Sjeng		&		55			&				&				&		46		&				&			\\
					& H264		&		54			&				&				&		47		&				&			\\	\hline
	\end{tabular}
\end{table}

Table~\ref{tab:BMUsed} lists the applications used in our study from different benchmarks. We have identified the applications under six domains.
Four of the six domains are coming from the famous \emph{MiBench} benchmark suite~\cite{Guthaus01MiBench} and they are \emph{Automotive (AM)}, \emph{Office (OF)}, 
\emph{Security (SE)} and \emph{Telecomm (TC)}.  The next domain contains four applications from \emph{MediaBench}~\cite{lee97MediaBench} benchmark suite
and the last domain is a set of integer applications from \emph{Spec2006} CPU~\cite{Henning06SPEC} benchmark suite. 

\section{Results and Analysis}\label{results}
In this section we present the results we obtained from our extensive instruction-level study of
reusability and extra cost of  the mASIPs with a carefully selected set of target  applications.
Reusability of instructions, the $Reusability\,Factor$, is defined as in Equation~\ref{eq1} using
the \emph{base instructions} and extra cost of supporting an application/domain, the $Extra\,Cost\,Factor$,
is defined using Equation~\ref{eq2}. We further analyse the results in order to identify suitable
patterns  and  behaviours  that  could be used in building a multi-application based ASIP design
automation tool.

In Table~\ref{tab:IntraDomain}, we show the instruction set selection for intra-domain target application
set. That is, a particular target mASIP can only be deployed with applications from a single domain. Therefore,
we have repeated the experiment six times, one for each domain targeted and the results are reported. It is 
worth to note that the instruction set selection for these experiments are groups separately for \emph{ARM-Thumb} (columns 3-5)
and \emph{PISA} (columns 6-8). Columns 3,4, and 5 gives the number of instructions of individual application,
intersection of all four applications of the particular domain, and the union of four applications of the
same domain in case of \emph{ARM-Thumb} ISA. Similar results are reported for \emph{PISA} in columns 6,7, and 8.

Let us take one of these six experiments of the \emph{ARM-Thumb} ISA, \emph{Automotive} of Table~\ref{tab:IntraDomain}, which 
contains the following applications: \emph{BasicMath}, \emph{BitCount}, \emph{QSort}
and \emph{Susan}. For this experiment, the target applications are the four mentioned earlier and therefore
our mASIP can support one or many of the four applications. Therefore, we have computed the intersection of 
the instruction sets from these four applications as our base instruction set and this number is 23. Now, if we 
are to deploy \emph{BasicMath} (which is having a total of 33 instructions as shown in column 3 of  Table~\ref{tab:IntraDomain}) 
on top of our base instruction set, we need to include 10 more instructions. Similar numbers for \emph{BitCount}, \emph{QSort} 
and \emph{Susan} are 23, 2 and 22 respectively. In addition, if we are to deploy all four applications at the same time, 
the total number of instructions required are 49 including the base instruction set, this is given in column 5 as the union value. 
The rest of the figures  in Table~\ref{tab:IntraDomain} are  similar results for the
experiments conducted in the rest of the domains namely \emph{Office}, \emph{Security}, \emph{Telecomm}, \emph{MediaBench}
and \emph{Spec2006}.

\begin{table}[ht!]
	\scriptsize
	\centering
	\caption{Instruction-set Selection for Inter-Domain Applications}
	\label{tab:InterDomain}
	\begin{tabular}{|l|l|c|c|c|c|c|c|} \hline
				    & 			& \multicolumn{6}{|c|}{\bf Number of Instructions} \\ \cline{3-8}
	{\bf Group} 	& {\bf Domain}	& \multicolumn{3}{|c|}{\bf ARM-Thumb ISA}& \multicolumn{3}{|c|}{\bf PISA}\\ \cline{3-8} 
					&			&	Indiv.	 		& 	Inter.		& Union 			&	Indiv.		& Inter.			& Union		\\	\hline
					& 	AM		&		45 			&				&				&	37			&				&			\\
	SET-01		 	& 	OF		&		51			&		14		& 	54			&	48			&	12			&	48		\\
					& 	MB		&		53			&				&				&	46			&				&			\\
					& 	SE		&		54			&				&				&	48			&				&			\\	\hline
					
					& 	AM		&		45 			&				&				&	37			&				&			\\
	SET-02		 	& 	OF		&		51			&		15		& 	55			&	48			&	13			&	48		\\
					& 	MB		&		53			&				&				&	46			&				&			\\
					& 	SP		&		55			&				&				&	48			&				&			\\	\hline

					& 	AM		&		45 			&				&				&	37			&				&			\\
	SET-03		 	& 	OF		&		51			&		14		& 	53			&	48			&	11			&	48		\\
					& 	MB		&		53			&				&				&	46			&				&			\\
					& 	TC		&		52			&				&				&	42			&				&			\\	\hline

					& 	AM		&		45 			&				&				&	37			&				&			\\
	SET-04			& 	OF		&		51			&   		14		& 	55			&	48			&	12			&	48		\\
					& 	SE		&		54			&				&				&	48			&				&			\\
					& 	SP		&		55			&				&				&	48			&				&			\\	\hline

					& 	AM		&		45			&				&				&	37			&				&			\\
	SET-05		 	& 	OF		&		51			&		13		& 	54			&	48			&	11			&	48		\\
					& 	SE		&		54			&				&				&	48			&				&			\\
					& 	TC		&		52			&				&				&	42			&				&			\\	\hline

					& 	AM		&		45 			&				&				&	37			&				&			\\
	SET-06		 	& 	OF		&		51			&		14		& 	55			&	48			&	11			&	48		\\
					& 	SP		&		55			&				&				&	48			&				&			\\
					& 	TC		&		52			&				&				&	42			&				&			\\	\hline

					& 	AM		&		45 			&				&				&	37			&				&			\\
	SET-07		 	& 	MB		&		53			&		16		& 	55			&	46			&	12			&	48		\\
					& 	SE		&		54			&				&				&	48			&				&			\\
					& 	SP		&		55			&				&				&	48			&				&			\\	\hline
					
					& 	AM		&		45 			&				&				&	37			&				&			\\
	SET-08		 	& 	MB		&		53			&		14		& 	54			&	46			&	11			&	48		\\
					& 	SE		&		54			&				&				&	48			&				&			\\
					& 	TC		&		52			&				&				&	42			&				&			\\	\hline

					& 	AM		&		45 			&				&				&	37			&				&			\\
	SET-09		 	& 	MB		&		53			&		17		& 	55			&	46			&	11			&	48		\\
					& 	SP		&		55			&				&				&	48			&				&			\\
					& 	TC		&		52			&				&				&	42			&				&			\\	\hline

					& 	AM		&		45 			&				&				&	37			&				&			\\
	SET-10			& 	SE		&		54			&   		14		& 	55			&	48			&	11			&	48		\\
					& 	SP		&		55			&				&				&	48			&				&			\\
					& 	TC		&		52			&				&				&	42			&				&			\\	\hline

					& 	OF		&		51			&				&				&	48			&				&			\\
	SET-11		 	& 	MB		&		53			&		18		& 	55			&	46			&	15			&	48		\\
					& 	SE		&		54			&				&				&	48			&				&			\\
					& 	SP		&		55			&				&				&	48			&				&			\\	\hline

					& 	OF		&		51 			&				&				&	48			&				&			\\
	SET-12		 	& 	MB		&		53			&		16		& 	54			&	46			&	13			&	48		\\
					& 	SE		&		54			&				&				&	48			&				&			\\
					& 	TC		&		52			&				&				&	42			&				&			\\	\hline

					& 	OF		&		51 			&				&				&	48			&				&			\\
	SET-13		 	& 	MB		&		53			&		17		& 	55			&	46			&	12			&	48		\\
					& 	SP		&		55			&				&				&	48			&				&			\\
					& 	TC		&		52			&				&				&	42			&				&			\\	\hline
					
					& 	OF		&		51 			&				&				&	48			&				&			\\
	SET-14		 	& 	SE		&		54			&		16		& 	55			&	48			&	12			&	48		\\
					& 	SP		&		55			&				&				&	48			&				&			\\
					& 	TC		&		52			&				&				&	42			&				&			\\	\hline

					& 	MB		&		53 			&				&				&	46			&				&			\\
	SET-15		 	& 	SE		&		54			&		19		& 	55			&	48			&	13			&	48		\\
					& 	SP		&		55			&				&				&	48			&				&			\\
					& 	TC		&		52			&				&				&	42			&				&			\\	\hline
	\end{tabular}
\end{table}

From the values in Table~\ref{tab:IntraDomain}, the number of \emph{base instructions} as a percentage to the union,
total number of instructions in the mASIP (the $Reusability\,Factor$ as per Equation~\ref{eq1}) are calculated and
are: 47\%, 49\%, 53\%, 45\%, 80\% and 78\% for \emph{Automotive}, \emph{Office}, \emph{Security}, \emph{Telecomm}, \emph{MediaBench}
and \emph{Spec2006} respectively. The average (arithmetic mean) of these numbers is 59\% and can be considered
as our mean $Reusability\,Factor$ for the six experiments we conducted for \emph{ARM-Thumb}.
Using the values in the same table, the numbers of instructions required to deploy a particular application
on a mASIP in addition to the \emph{base instruction} set of the domain as a percentage to the union 
(the $Extra\,Cost\,Factor$ as per Equation~\ref{eq2}) are calculated and the average values for each 
domain are: 29\%, 30\%, 27\%, 32\%, 13\%, and  13\% for \emph{Automotive}, \emph{Office}, \emph{Security}, \emph{Telecomm}, \emph{MediaBench}
and \emph{Spec2006} respectively. 
The average (arithmetic mean) of these numbers is 24\% and can be considered as the mean $Extra\,Cost\,Factor$ 
for the six domains for \emph{ARM-Thumb}. Given that the experiments are for intra-domain applications, the reusability is expected to be higher
than the extra cost (i.e., applications in an intra-domain should have very similar functionalities hence would require a 
similar set of instructions) which can be verified from the numbers we obtained here.
From Table~\ref{tab:IntraDomain} we obtained similar results for \emph{PISA}. The mean $Reusability\,Factor$ for
the six experiments we conducted for \emph{PISA} is 49\% and the mean $Extra\,Cost\,Factor$ for the six domains for
\emph{PISA} is 26\% proving our expectation that reusability is higher than the extra cost in intra-domain
applications.


In Table~\ref{tab:InterDomain}, we tabulate the instruction set selection for inter-domain target application
sets. That is, a particular target mASIP can be deployed with applications from different application
domains. We assume that a mASIP can integrate at most four application domains (the rest of the combinations
are not reported due to lack of space) and given that we have six domains, we could have
15 combinations ($^{6}C_{4}$). As shown in Table~\ref{tab:InterDomain}, we named each of this combination a SET
and therefore we have repeated the experiment 15 times, one for each SET and the results are reported. Columns 3, 4, and 5 show
the number of instructions for each domain, the intersection, and the union of all four domains of the particular set 
respectively in case of \emph{ARM-Thumb} ISA.  Columns 6-8 show similar results for \emph{PISA}.

Let us consider the numbers in table~\ref{tab:InterDomain}. 
The domains taken into the set (\emph{SET-01}) are \emph{Automotive (AM)}, 
\emph{Office (OF)}, \emph{MediaBench (MB)} and \emph{Security (SE)} which contain a total of 16 applications, four each.
For this experiment, the target applications are the 16 mentioned earlier and therefore
our mASIP can support one or many of the sixteen. Therefore, in case of \emph{ARM-Thumb} ISA, we have computed the intersection of 
the instruction sets from these 16 applications as our base instruction set and this number is 14. 
Now, if we are to deploy all the applications in \emph{Automotive} on top of our \emph{base instruction} set,
we need to include 31 more instructions as shown in Table~\ref{tab:InterDomain}. 
Similar numbers for \emph{Office (OF)}, \emph{MediaBench (MB)} and \emph{Security (SE)} are 37, 39 and 40
respectively. In addition, if we are to deploy all 16 applications at the same time, the total number of instructions
required are 54 including the \emph{base instruction} set. The rest of the figures in Table~\ref{tab:InterDomain} 
are for the experiments conducted for the rest of the 14 sets of \emph{ARM-Thumb} and 15 sets of \emph{PISA}.


From the numbers in Table~\ref{tab:InterDomain}, in case of inter domain \emph{ARM-Thumb} experiments the
mean $Reusability\,Factor$ is 28\% and the mean $Extra\,Cost\,Factor$ is 67\%. Given that the experiments are for inter-domain applications,
the $Reusability\,Factor$ is expected to be lower than the $Extra\,Cost\,Factor$ (i.e., applications from different domains will
have quite varying functionalities, hence contain different instructions) which is reflected in our experiments.

\begin{figure*}[ht!]
\centering
\includegraphics[width=15cm]{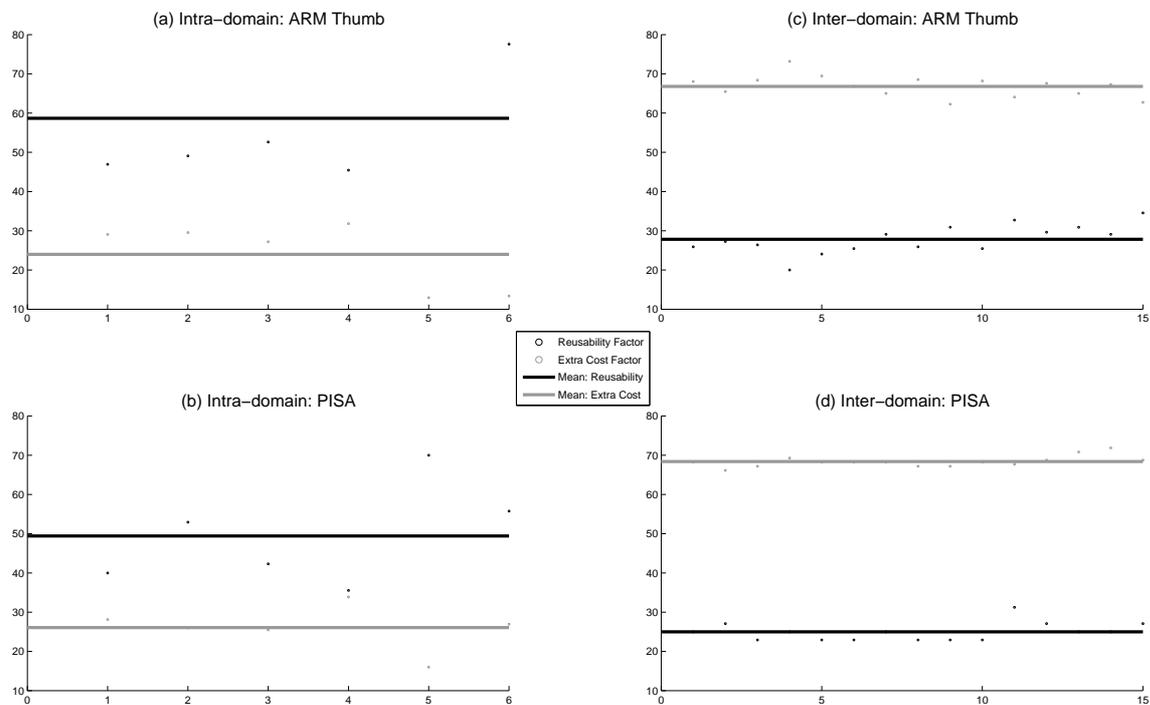}
\caption{Reusability and Extra Cost Factors for Intra- and Inter-Domain Applications}
\label{fig:resultsplot}
\end{figure*}

By using the values shown in table~\ref{tab:InterDomain} similar results are calculated in case of \emph{PISA}. The
mean $Reusability\,Factor$ is 25\% and the mean $Extra\,Cost\,Factor$ is 68\% still proving our expectations that the 
reusability is lower than the extra cost for the inter-domain scenario, even for a different architecture.


In Figure~\ref{fig:resultsplot} the four graphs are depicting the reusability factor values and extra cost factor values we
have discussed previously in Intra-domain and Inter-domain  experiments for \emph{ARM-Thumb} and \emph{PISA} target architectures. 
In Figure~\ref{fig:resultsplot} (a), (b) representing intra-domain experiments the mean of reusability factor is higher 
than the mean of extra cost factor whereas, in Figure~\ref{fig:resultsplot} (c), (d)
representing inter-domain experiments the mean of extra cost factor is higher than the mean of reusability factor. 


\section{Conclusion} \label{conclusion}
We  performed  an  extensive study in the instructions for a  multi-application
based ASIP, which was meant to  execute  inter-domain and intra-domain applications. \emph{MiBench}, \emph{MediaBench} and \emph{SPEC2006} benchmarks are experimented for \emph{ARM-Thumb}
and \emph{PISA} architectures. Our experiments prove that the \emph{reusability} in instructions is larger than the 
\emph{extra cost} for intra-domain and smaller for inter-domain applications. This justifies our hypothesis that instruction-level analysis
is useful to design multi-application based ASIPs, regardless of the instruction set architecture.  
 

\bibliographystyle{abbrv}
\bibliography{paperref_reduced}
\end{document}